\documentclass[10pt]{article}
\usepackage{amssymb,amsmath,amsfonts}
\usepackage{graphicx,cite}

\title{Bohm's quantum ``non-mechanics'':\\ An alternative quantum theory\\ with its own ontology?}

\author{\'Angel S. Sanz \\ \\
Department of Optics, Faculty of Physical Sciences,\\ Universidad Complutense de Madrid\\
Pza.\ Ciencias 1, Ciudad Universitaria -- 28040 Madrid, Spain}

\date{}

\begin{document}

\maketitle

\begin{abstract}
L'aspect ontologique de la m\'ecanique bohmienne, en tant que th\'eorie des variables
cach\'ees qui nous fournit une description objective d'un monde quantique sans observateurs,
est largement connu.
Pourtant, son caract\`{e}re pratique est de plus en plus accept\'e et reconnu, car il
s'est av\'er\'e \^{e}tre une ressource efficace et utile pour aborder, explorer, d\'ecrire
et expliquer de tels ph\'enom\`{e}nes.
Cet aspect pratique \'emerge pr\'ecis\'ement lorsque l'application pragmatique du formalisme
l'emporte sur toute autre question d'interpr\'etation, encore sujet \`a d\'ebat et \`a
controverse.
\`{A} cet \'egard, notre objectif est de montrer et de discuter ici comment la m\'ecanique bohmienne
met en valeur de mani\`{e}re naturelle une s\'erie de caract\'eristiques dynamiques difficiles
\`{a} d\'ecouvrir \`{a} travers d'autres approches quantiques.
Cela vient du fait que la m\'ecanique bohmienne permet d'\'etablir un lien direct entre
la dynamique des syst\`{e}mes quantiques et les variations locales de la phase quantique
associ\'ees \`{a} leur \'etat.
Pour illustrer ces faits, deux mod\`{e}les simples de ph\'enom\`{e}nes quantiques
physiquement \'eclairants ont \'et\'e choisis, \`{a} savoir la dispersion d'un paquet
d'ondes gaussiennes libres et l'interf\'erence \`{a} deux fentes de type Young.
Comme il est montr\'e ici, les r\'esultats de leur analyse offrent une compr\'ehension
alternative de la dynamique affich\'ee par ces ph\'enom\`{e}nes quantiques en termes du champ de
vitesse local sous-jacent, qui relie la densit\'e de probabilit\'e au flux quantique.
Ce champ, qui n' exprime rien d' autre que la condition de guidage en m\'ecanique bohmienne standard,
acquiert ainsi un r\^{o}le de premier plan pour comprendre la dynamique quantique, en tant
que m\'ecanisme responsable de cette dynamique.
Cela va au-del\`{a} du r\^{o}le passif g\'en\'eralement attribu\'e au champ de la vitesse 
locale en m\'ecanique bohmienne, o\`{u} traditionnellement l' on accorde plus d' attention aux trajectoires et au potentiel quantique.

\vskip 1cm

{\it
The ontological aspect of Bohmian mechanics, as a hidden-variable
theory that provides us with an objective description of a quantum world without
observers, is widely known.
Yet its practicality is getting more and more acceptance and relevance, for it has proven
to be an efficient and useful resource to tackle, explore, describe and explain such
phenomena.
This practical aspect emerges precisely when the pragmatic application of the formalism
prevails over any other interpretational question, still a matter of debate and controversy.
In this regard, the purpose here is to show and discuss how Bohmian mechanics emphasizes
in a natural manner a series of dynamical features difficult to find out through other
quantum approaches.
This arises from the fact that Bohmian mechanics allows us to establish a direct
link between the dynamics exhibited by quantum systems and the local variations of the
quantum phase associated with their state.
To illustrate these facts, simple models of two physically insightful quantum phenomena
have been chosen, namely, the dispersion of a free Gaussian wave packet and Young-type
two-slit interference.
As it is shown, the outcomes from their analysis render a novel, alternative understanding of
the dynamics displayed by these quantum phenomena in terms of the underlying local velocity
field that connects the probability density with the quantum flux.
This field, nothing but the so-called guidance condition in standard Bohmian mechanics, thus
acquires a prominent role to understand quantum dynamics, as the mechanism responsible for
such dynamics.
This goes beyond the passive role typically assigned to this field in Bohmian mechanics,
where traditionally trajectories and quantum potentials have received more attention
instead.}
\end{abstract}



\section{Introduction}
\label{sec1}

The quantum approach that is commonly known as Bohmian mechanics\footnote{Within the field
of the quantum foundations, Bohmian mechanics is also widely known as the de Broglie-Bohm
interpretation.
However, in recent times the term Bohmian mechanics has become more widespread when it is
used in applications.
This will be the term also considered here.}
has been a source of controversy since its inception \cite{bohm:PR:1952-1,bohm:PR:1952-2},
formerly intended as a simple counter-proof to Von Neumann's theorem \cite{vonNeumann-bk:1932}
on the incompatibility between quantum mechanics and any possibility to complete this theory
with the introduction of local hidden variables.
Therefore, after having worked for a long time taking Bohmian mechanics as a
fundamental theoretic-analytical tool to explore, understand and describe different aspects
of quantum and optical phenomena, one learns to live with a series of recurrent questions
from colleagues and reviewers: Why should anyone be interested in Bohmian-related ``stuff''?
Which new physics does Bohmian mechanics add with respect to the other more conventional
quantum approaches?
Is it not redundant?
In addition, if the always appealing though controversial concept of hidden variable also
appears without having made explicit mention to it (or without having mentioned it at all),
things become even a bit more complicated.

All in all, the general trend seems to be smoothly changing towards what could be considered
to be, say, a more Bohmian-friendly attitude than it was ten or twenty years ago (not to say
earlier on).
Since the 1990s an increasing number of monographs have been published on the issue
\cite{holland-bk,bohm-hiley-bk,cushing-bk:1996,wyatt-bk,bacciagaluppi-valentini-bk,duerr-bk:2009,duerr-bk:2013,hughes-bk,chattaraj-bk,oriols-bk,sanz-bk-1,sanz-bk-2,bricmont-bk}.
These works describe and discuss the physical (and metaphysical) implications of Bohmian
mechanics, revisit the standard quantum formulation in terms of this approach or provide a
detailed account on its applications to different physical problems, which the interested
reader is kindly invited to consult (of course, bearing in mind that the list of works is far
from being complete, yet it serves to the purpose of illustration).
When facing such a flourishing landscape of new developments in the field, one feels compelled
to revisit the above questions, particularly the one about why any attention should be paid at
all to the Bohmian approach beyond the hidden-variable issue, i.e., beyond ontological questions
related to the completeness of quantum mechanics.

After a long and tough way, some conclusions have come up in that regard,
partly collected and discussed in previous works \cite{sanz:AJP:2012,sanz:foundphys:2015,sanz:FrontPhys:2019}.
Now, getting back to Bell's pedagogical view on Bohm's mechanics \cite{bell-bk},
the very first point that one should address is whether, keeping our feet on solid ground,
beyond metaphysical questions, this approach provides us with a natural scenario to think
the physics of quantum phenomena.
This does not mean to consider that Bohm's particle trajectory is the actual trajectory
followed by a real quantum particle\footnote{Note that the concept of trajectory needs not
be necessarily associated with the actual position of a particle or that of its center of mass.
Rather, it should be understood in a broad sense, i.e., as describing the evolution in time
of any type of degree of freedom (vibrations, rotations, etc.), which is a point often
neglected in discussions around Bohm's theory, where trajectories are immediately and uniquely
related to point-like (structureless) particles.
In this sense, Bohmian mechanics transcends the oversimplified framework that associates
the approach with a theory of motion for quantum particles; it can be applied to any aspect
of matter that is accounted for by Schr\"odinger's equation, although providing the
corresponding trajectories with the appropriate interpretation (i.e., in compliance with
the context considered).}.
Yet, the possibility to introduce this ``forbidden'' element in quantum mechanics allows us
to understand the evolution of quantum waves on formal and conceptual grounds analogous to
those used to describe the evolution of classical action in phase space, namely, the theory
of characteristics \cite{courant-hilbert-bk-2} and dynamical flows \cite{jordan-bk:1999}.
Of course, there are certain formal subtleties that generate necessary differences between
the classical and the quantum descriptions:
while space point dynamics is well-defined in the former, the latter precludes it in the same
terms, because it assumes mutual spatial coherence among different spatial points (nicely
evidenced by the Moyal-Wigner representation), which in turn implies a revision of the
laws of motion.
Nonetheless, this does not invalidate the existence of a common formal structure.

Stepping down from the formal level to, say, the level of our everyday experience, based on
real experiments with real quantum particles (including photons, whatever they might be),
several facts are worth noticing:
\begin{itemize}
 \item[i)] The evolution of quantum particles takes place in real time.
 Quantum particles cannot (or should not) be dissociated from the reality we live in (and
 where they also live in).
 In a typical diffraction or scattering experiment, for instance, pushing a trigger on they
 start being launched; pushing the trigger off the flux of particles ends.
 Now, in the meantime, each one of such particles has moved from wherever they were at a
 $t_0$ to somewhere else at $t > t_0$ (relativistic issues are left aside for simplicity
 and because they are not necessary at all in the discussion).

\item[ii)] The quantum theory is a statistical theory, where the so-called observables
 correspond to statistical quantities.
 Single events or realizations, e.g., the detection of one particle at a time $t$, do not
 provide us with any relevant information about the process investigated; to obtain precise
 (physically meaningful) information, a large number of events or realizations (detected
 particles) is required, which involves a statistical analysis.
 The probability distributions rendered by conventional quantum mechanics are directly
 related to this large-numbers approach, which is the way how we can access and investigate
 experimentally quantum systems.
 In this regard, it is worth noting that the widespread conception that a full interference
 pattern, for instance, is related with each single quantum particle is based on the
 experimental performance previous to the advent of quantum mechanics, as it is inferred
 from works at that time \cite{taylor:PCPS:1909,dempster:PhysRev:1927}.
 After all, the atomistic understanding of matter was not solidly grounded.
 But, more importantly, this view did not allow the settlement of solid conclusion about
 the inherent statistical nature of quantum particles.
 Only by the end of the 1970s and along the 1980s the refinement reached in the experimental
 techniques enabled single-particle production/detection
 \cite{davis:IEEEJQuantumElectron:1979,aspect:EPL:1986,kimura:OptCommun:1989},
 which in turn gave complete sense to the statistical meaning of quantum observables, as
 related to a collection of individually (detected) events.
 This leads us directly to fact (iii) below.

\item[iii)] Even when it can be experimentally shown that there is no time-correlation
 between the evolution of one particle ``identically'' prepared with respect to another
 particle that precedes it, the two particles behave as if they shared some kind of
 fundamental information.
 Independently of their nature, all quantum particles exhibit the same behavior in
 event-by-event experiments (photons \cite{weis:AJP:2008,weis:EJP:2010,padgett:AJP:2016},
 electrons \cite{pozzi:AJP:1973,pozzi:AJP:1976,pozzi:AJP:2007,pozzi:EJP:2013,tonomura:ajp:1989,batelaan:NJP:2013},
 atoms \cite{shimizu:pra:1992} or large macromolecular complexes
 \cite{arndt:Nature:1999,arndt:NatCommun:2011,arndt:NNanotech:2012}).
 This behavior has also been observed even in classical-type processes\footnote{Here, the
 notion of ``classical-type'' applied to quantum or optical processes will be understood in
 the context of wave descriptions that include a partial or even total lack of coherence (and
 hence they are unable to display interference), regardless of how the latter arises.},
 such as imaging produced by objects under extremely faint illumination conditions.
 In these cases, the image becomes apparent after a rather long exposure time, once
 the number of collected photons is relatively high, as it is shown and discussed in
 \cite{rose:AdvBiolMedPhys:1957}, in the context of the limitations imposed on vision
 by the quantum (granular) nature of light.
 This is exactly the same problem that affect (the imaging of) interference patterns when
 the source is very weak under total coherence conditions.
\end{itemize}

The above facts are relatively well known and hence they might seem natural to the reader
(even trivial).
However, they pave the way for event-to-event statistical descriptions of quantum systems
to the detriment of the single-particle approaches typically associated with Schr\"odinger's
equation.
It is precisely here where Bohmian mechanics comes into play: it gathers all the formal
elements to be consistent with quantum mechanics (it is actually quantum mechanics) without
the necessity to introduce any additional quantities or approximations.
Indeed it is an ideal candidate to investigate quantum dynamics in conformity with the above
three experimental facts: evolution in real time, observables arising from statistics
over individual events and uncorrelated realizations (events).
Leaving aside complications arising from computational implementations, we now have a
convenient tool to compare on equal (statistical) footing experiment and theory (detections
vs realizations).
This is precisely a legitimate argument to respond the everlasting criticism on the
additional physical content, which also leaves aside the hidden-variable issue,
because both concepts and formalism are well defined.
In fact, the role of the so-called quantum postulates is diminished.
So, what else could one wish?

Thus, so far, it is clear that, provided all sources of controversy are left aside
(at least in line with the renowned Copenhagian `shut up and calculate!'
\cite{mermin:PhysToday:1989}), nothing wrong is found in Bohmian mechanics, nor even
one needs to give further explanations on which new physics it provides us with.
Of course, there are some subtleties that make Bohmian mechanics different at an intuitive
level from the point of view of classical Newtonian mechanics, but this is also legitimate
for, after all, quantum mechanics itself is conceptually different from classical mechanics,
as stressed by the Bell inequalities \cite{bell:physics:1964,bell:RMP:1966}.
As mentioned by Hiley \cite{muser-hiley:SciAm:2013}, this difference used to be remarked by
Bohm by talking about Bohmian `non-mechanics', since this quantum approach {\it sensu
stricto} has little in common with mechanics.
Note, for instance, that the standard concept of force gets diluted
with the introduction of a quantum force mediated by Bohm's quantum potential.
Yet this is a hydrodynamic-like model that serves to the purpose, making more apparent
dynamical behaviors that go beyond our classical intuition, although they rule nature at
the microscopic and mesoscopic levels (with important implications on the macroscopic one).

Following the preceding discussion, the purpose here is to show and discuss a
``non-mechanical'' perspective of the Bohmian approach, that is, avoiding the traditional
ideas of quantum potential and quantum force, and trying to ground the description of quantum
phenomena on a ``non-observable'' (in the standard sense), namely, the quantum phase field
associated with the system state.
Typically, Bohmian mechanics includes discussions that turn around the concepts of Bohm's
quantum potential (or the forces generated by this potential) and how it rules the behavior
of the so-called Bohmian trajectories.
However, the quantum potential is only a measure of the curvature of the probability density
and, therefore, one feels compelled to find other alternative mechanisms responsible for the
dynamics exhibited by quantum systems.
The quantum phase and, more specifically, its local variations, though, have not been much
exploited in the literature, although they translate into a local velocity field that, in
principle, can be measured by means of the so-called weak measurements
\cite{aharonov:PRL:1988,sudarshan:PRD:1989,wiseman:NewJPhys:2007}.
When this velocity field, which is denoted as ``local'' because the system flow is determined
by its local value, is considered, a series of interesting properties emerge, which are not
proper of Bohmian mechanics, but of quantum mechanics in general, although they cannot be
easily perceived with other quantum formulations.
For instance, the so-called non-crossing rule in Bohmian mechanics is nothing but a
combination of the single-valuedness of the quantum phase (except for integer $2\pi$-jumps,
which are unnoticeable in the velocity field) and the dynamical domains determined by the
velocity field.
In order to illustrate these properties, a simple model of Gaussian diffraction and Young-type
interference are going to be analyzed in next sections, because of their interest not only
in quantum mechanics, but also in wave optics, which shares common theoretical grounds with
the former (despite the latter is typically regarded as a ``classical'' theory).

The work has thus been organized as follows.
Section~\ref{sec2} introduces and discusses some fundamental aspects of
Bohmian mechanics in the direction pointed out above, making emphasis on those formal aspects
that put the approach at the level of any other quantum representation rather than in those
that have traditionally associated it with a theory without observers, where trajectories are
(unfoundedly) related to paths followed by quantum particles.
Furthermore, on the analytical level, some particular aspects of the quantum potential are
illustrated by employing a simple diffraction model consisting of a free Gaussian wave
packet.
Section~\ref{sec3} is devoted to revisit and discuss some physical consequences related
to a Young-type interference.
To support the interest in the theory, particularly taking into account the discrete nature
of quantum phenomena, first the outcomes from a simple event-by-event Young-type experiment
are reported and discussed.
Then, Young-type fringes are analytically described in terms of a simple model consisting of
a coherent superposition of two Gaussian wave packets.
This models describes in a convenient manner the emergence of interference fringes along the
transverse direction (assuming the matter wave propagates forward with a fast speed, as it
is usually the case in slit and grating diffraction experiments).
More specifically, it will serve to show the main difference between the physics linked to
the quantum potential and the physics rendered by the velocity field.
Finally, the work concludes with a series of remarks summed up in Sec.~\ref{sec4}.


\section{A critical view on Bohmian mechanics: Concepts and formalism}
\label{sec2}


\subsection{Hidden variables {\em vs} experimental facts}
\label{sec21}

Since much has already been said in the literature about conceptual and formal aspects of
Bohmian mechanics, this section will be devoted to highlight other formal aspects, which
have not been so extensively considered.
Yet these aspects provide us with a different perspective of both Bohmian mechanics
itself and the quantum phenomena in general.
They will also be useful to get a better and broader understanding of the discussion in
Sec.~\ref{sec3}, at the same time that they in compliance with the statement made by Bohm
regarding his reformulation of quantum mechanics \cite{bohm:PR:1952-1}:
%
\begin{quotation}
\noindent
``[\ldots]\ the suggested interpretation provides a broader conceptual framework than the
usual interpretation, because it makes possible a precise and continuous description of all
processes, even at the quantum level.
This broader conceptual framework allows more general mathematical formulations of
the theory than those allowed by the usual interpretation.''
\end{quotation}

The above statement refers to interpretation, that is, how we should or could consider that
real particles behave in space and time.
Now, to put forth the question on a real-life context, consider the chip of a CCD
made of an array of pixels and connected to a screen where the detection of a photon in a
pixel translates into a scintillation on the screen.
Is there any good or deep reason preventing us from joining the scintillation (single photon
detection) with a specific source point at a previous time?
That is, can we establish a causal connection between the two points?
In principle, it seems there is no empirical evidence neither in favor nor against it.
However, assuming that we accept that such a link can be established, the next question is
whether the connection can be done by means of a smooth trajectory, more specifically, a
Bohmian trajectory.
This has been a central question in Bohmian mechanics since its beginning in the early 1950s.
Although the Bohmian approach prescribes a precise way to proceed, we have no way to
demonstrate that nature operates the same way; other alternative approaches could also be
formulated with a similar result, but without the need to consider a smooth causal connection.
This is the case, for instance, of the stochastic approaches proposed by Bohm and Vigier
\cite{bohm:pr:1954} (later on also considered by Bohm and Hiley \cite{bohm:PhysRep:1989}) or
by Nelson \cite{nelson:pr:1966}.
Nonetheless, it is clear that there is an appealing feature in Bohmian trajectories over
other types of trajectory-based approaches: it renders a fair reproduction of the detection
process, statistically speaking, at the same time that offers a precise description of the
system evolution (spatial diffusion, diffractive effects or whirlpool-type motion).


\subsection{Equations of motion and trajectories}
\label{sec22}

The standard starting point of Bohmian mechanics consists in recasting
Schr\"odinger's equation in the form of two coupled real-valued partial differential
equations \cite{bohm:PR:1952-1,holland-bk}.
This is achieved by writing the wave function (formerly given in the configuration
representation) in polar form, as
\begin{equation}
 \Psi({\bf r},t)= A ({\bf r},t) e^{iS({\bf r},t)/\hbar} .
 \label{eq1}
\end{equation}
This nonlinear transformation allows us to pass from the complex field variable
$\Psi$ to two real field variables, namely, an amplitude $A$ and a phase $S$.
This ansatz was formerly considered by Dirac \cite{dirac:PRSLA:1931}, in connection with the
existence of quantized singularities (magnetic monopoles), and by Pauli \cite{pauli-hbk-1},
in the context of quantum-classical correspondence.
After substitution into the time-dependent Schr\"odinger equation,
\begin{equation}
 i\hbar\ \frac{\partial \Psi}{\partial t} = - \frac{\hbar^2}{2m} \nabla^2 \Psi + V \Psi ,
 \label{eq2}
\end{equation}
and then proceeding with some simple algebraic manipulations, the real and imaginary parts
of the resulting equation give rise to two coupled partial differential equations:
\begin{subequations}
\begin{eqnarray}
 \frac{\partial A^2}{\partial t}\ & +\ & \nabla \cdot \left( A^2 \frac{\nabla S}{m} \right) = 0 ,
 \label{eq3a} \\
 \frac{\partial S}{\partial t}\ & +\ & \frac{(\nabla S)^2}{2m} + V - \frac{\hbar^2}{2m}\frac{\nabla^2 A}{A} = 0 .
 \label{eq3b}
\end{eqnarray}
 \label{eq3}
\end{subequations}

The first equation, Eq.~(\ref{eq3a}), arising from the imaginary part of the Schr\"odinger
equation, deals with the spatial diffusion or dispersion of the probability density,
$\rho ({\bf r},t) = A^2 ({\bf r},t)$.
This is a continuity equation relating the evolution of the probability density in a
position (configuration) space with the vector quantity
\begin{equation}
 {\bf J}({\bf r},t) = A^2 ({\bf r},t) \ \frac{\nabla S ({\bf r},t)}{m}
  = \frac{1}{m}\ {\rm Re} \left\{ \Psi^* ({\bf r},t) \hat{\bf p} \Psi ({\bf r},t) \right\} ,
 \label{eq4}
\end{equation}
where $\hat{\bf p} = - i\hbar \nabla$ is the expression of the momentum operator in the
configuration representation.
Equation~(\ref{eq4}) describes the quantum flux or current density, a well-known quantity
in quantum mechanics \cite{bohm-bk:QTh,schiff-bk}, introduced at the very beginning
of any elementary course on the subject.
As it can be noticed on the r.h.s.\ of the first equality in Eq.~(\ref{eq4}), the quantum
flux can be rewritten in a more compact form as
\begin{equation}
 {\bf J}({\bf r},t) = \rho({\bf r},t) {\bf v}({\bf r},t) .
 \label{eq4b}
\end{equation}
This expression not only makes emphasis on the causal relationship between the probability
density $\rho$ and its dispersion in terms of the quantum flux, but it also makes more
apparent the mechanism for such a dispersion, namely, the presence of an underlying
local velocity field,
\begin{equation}
 {\bf v}({\bf r},t) = \frac{{\bf J}({\bf r},t)}{\rho({\bf r},t)}
  = \frac{\nabla S ({\bf r},t)}{m} .
 \label{eq5b}
\end{equation}
Physically, this vector field accounts for the density flow rate through the point ${\bf r}$
at a time $t$, i.e., its value changes locally following the variations of the quantum phase
$S$, unlike the average drift value obtained from the expectation value of the momentum,
$\langle \hat{\bf p} \rangle/m$.

The velocity field (\ref{eq5b}) is not a proper quantum observable in spite of its
connection to the usual momentum operator $\hat{\bf p}$.
Yet, it is going to play a fundamental role in the quantum dynamics, because of the
information that it provides on the concentration, expansion, diversion or rotation of the
quantum flow at each point.
The natural question that arises here is why this quantity is not mentioned at all
in any standard course on quantum mechanics, although it is well defined and provides extra
local information on the deformation of the probability density in the configuration space.
Actually, not only in standard quantum mechanics, but also in Bohmian mechanics this quantity
is often neglected in favor of other quantities, such as the so-called Bohm's quantum
potential, usually required to explain the behavior displayed by Bohmian trajectories.

To understand the above statement, let us get back to Eqs.~(\ref{eq3}).
The second differential equation, Eq.~(\ref{eq3b}), encoded in the real part of
Schr\"odinger's equation, keeps a formal resemblance with the classical Hamilton-Jacobi
equation \cite{goldstein-bk}.
This classical Hamilton-Jacobi equation arises from the so-called Hamiltonian analogy
between mechanics and optics \cite{bornwolf-bk}, which establishes a connection (analogy)
between the wavefronts of optics (surfaces of constant phase) and surfaces of constant
mechanical action.
Hence, in the same way that light rays are perpendicular to the wavefronts at each point,
the Newtonian trajectories are perpendicular to constant-action surfaces.
Following this analogy, Bohm considered Eq.~(\ref{eq3b}) to be a quantum version of
the Hamilton-Jacobi equation, thus postulating the existence of a quantum Jacobi law of
motion \cite{bohm:PR:1952-1,holland-bk}
\begin{equation}
 \dot{\bf r} ({\bf r},t)= \frac{\nabla S ({\bf r},t)}{m} .
 \label{eq5}
\end{equation}
This equation of motion is known as the guidance equation.
In agreement with the widespread Bohmian interpretation for this equation, it rules the
(quantum) way how particles moves, since integrating in time with the corresponding initial
conditions one obtains swarms of (Bohmian) trajectories.
Extending this idea further beyond, one might conclude that particle instantaneous
positions (e.g., the trajectory of an electron or a photon in an interference experiment)
are actually described by the solutions to (\ref{eq5}), thus becoming a sort of
hidden causal variables.
However, as seen above, this equation of motion is exactly the same as Eq.~(\ref{eq5b}),
which naturally follows from the standard formulation (from the continuity equation),
without any need to postulate anything, nor even the existence of hidden variables.

In classical mechanics there is a clear and direct connection between the trajectories
obtained from Jacobi's law, which describe the dynamical properties displayed by a
phase-space distribution function, and Newtonian trajectories, which account for the
evolution of individual systems.
That is, there is a one-to-one correspondence between the descriptors for ensembles and
for individuals, which is ultimately based on experience (statistics is the large-particle
limit of dynamics).
However, the same connection cannot be established for quantum systems.
Note that quantum mechanics, which is a statistical theory itself (with some
peculiar properties that make it different from classical statistics), lacks a quantum
counterpart for individuals, thus avoiding us to compare the Bohmian trajectories that
describe the dynamical properties of the probability density (in configuration space)
with the dynamics exhibited by individual particle trajectories obtained from a
singe-body dynamical law, i.e., the quantum analog to Newton's trajectories.

Bohmian trajectories have long been identified with such quantum Newtonian trajectories.
However, this is not based on solid empirical grounds, but on a weak conceptual inference:
because the statement holds for classical particles, it must also hold for quantum ones.
This is a very weak argumentation, because classical mechanics
and classical statistical mechanics are based on different formal grounds (phase space)
than quantum mechanics (positions or momenta, but not both at the same time, unless we
pay a price for it, as it happens ih the Wigner-Moyal representation).
Establishing a strong unique connection would require empirical evidence beyond
statistics-based experiments and reconsidering theoretical models in the direction of
de Broglie's former ideas of wave fields and particles both coexisting but being different
physical entities \cite{broglie-bk:1960}, the stochastic causal model \cite{bohm:pr:1954,bohm:PhysRep:1989}
or subquantum Brownian-type models \cite{furth:ZPhys:1933,comisar:PhysRev:1965,nelson:pr:1966}.
This is an important point, for instance, when using Eq.~(\ref{eq5}) in the interpretation and
understanding of single-photon experiments \cite{kocsis:Science:2011,steinberg:SciAdv:2016},
since the information provided by such experiments is indeed understandable in terms of our
actual theories of light; trajectories inferred from the experiments thus do not represent
the actual motion of real photons, but just the average expansion or contraction undergone
by the wave field corresponding in the large photon-number limit.
In order to switch from matter waves to light in this regard, it can easily be noted that
Eq.~(\ref{eq5}) shows a certain reminiscence of the optical ray equation, which describes
rays as lines always perpendicular to the surfaces of constant phase and would arise from
the aforementioned Hamiltonian analogy \cite{bornwolf-bk} (which, in turn, underlies the
derivation of Schr\"odinger's equation).
This is precisely the same conclusion found by de Broglie earlier on\footnote{In this
regard, the interested reader might like to consult the work by Drezet and Stock
\cite{drezet:AnnFondLdBroglie:2021} on an original manuscript sent by Bohm to de Broglie
in 1951, which predates the renowned 1952 papers and, according to the authors, seems to
be its origin.}
\cite{broglie:CompRend-2:1926}, which in the case of classical light (which does
not follow Schr\"odinger's equation, but Maxwell's equations) works very nicely
\cite{prosser:ijtp:1976-1,sanz:AnnPhysPhoton:2010,sanz:JRLR:2010}.


\subsection{Bohm's quantum potential}
\label{sec23}

Unlike classical particles, the motion displayed by quantum particles\footnote{Following
the preceding discussion, the meaning of ``particle'' here is as denoted above,
i.e., as an entity that serves us to keep track of the quantum density flux.} that follow
Eq.~(\ref{eq5}) is affected not only by the forces induced by $V({\bf r},t)$, but also
by an additional term, as seen in Eq.~(\ref{eq3b}).
This is the so-called Bohm's quantum potential,
\begin{equation}
 Q({\bf r},t) = - \frac{\hbar^2}{2m} \frac{\nabla^2 A({\bf r},t)}{A({\bf r},t)}
   = - \frac{\hbar^2}{4m} \left\{ \frac{\nabla^2 \rho({\bf r},t)}{\rho({\bf r},t)}
     - \frac{1}{2} \left[ \frac{\nabla \rho({\bf r},t)}{\rho({\bf r},t)} \right]^2 \right\} .
 \label{eq6}
\end{equation}
This contribution to the quantum Hamilton-Jacobi equation has little in common with usual
potential functions acting on physical systems.
Rather it is associated with the local curvature of the amplitude of the wave function,
undergoing important values in those regions where the amplitude becomes negligible, but
not its Laplacian.
This happens in nodes and nodal lines, where the quantum force acting on the particle
becomes very intense and so the changes in ${\bf v}$.
However, this is all related to the quantum state of the system itself and not to any
external interaction.
This is more apparent if we look at the right-hand side of the second equality, which is
explicitly written in terms of $\rho$.
In a Bohmian sense, $\rho$ describes the statistical distribution of independent realizations,
i.e., it is produced by the cumulative effect arising after, for instance, launching a large
number of independent photons or electrons (but all subjected to the same experiment), and 
see how they start distributing spatially after a given (long enough) exposure time on the
corresponding detector \cite{tonomura:ajp:1989,weis:AJP:2008,padgett:AJP:2016,arndt:NNanotech:2012}.
How can independent realizations influence one another?

The above discussion leads us to a deeper question, namely, that of the reality of the wave
function as a physical field, beyond its usual conception as a probabilistic information
descriptor \cite{pusey:NaturePhys:2012}.
Quantum systems or, more strictly speaking, their seemingly random statistical distributions
would play the role of tracers that allow us to feel such a presence, in the same way, for
instance, that
iron powder allows us to make observable magnetic the line forces and, therefore, to detect
the presence of magnetic fields.
Of course, this does not provide any clue on the origin of this field or whether quantum
systems behave in the same way as Bohmian trajectories do (see discussion in next
section), but at least it seems there is a dynamical element that is totally neglected,
namely, the presence of an intrinsic velocity field.
This field does not require the presence (or existence) of a quantum potential, because it
is directly related to the phase (see below), hence removing the redundancy of describing
quantum effects in terms of $\rho$ and its curvature.
Furthermore, this quantity tells us that there is an underlying
statistical stream behavior not necessarily related with a quantum observable (although
its effects manifest through the topology displayed by $\rho$).

%
%

There is another important question regarding the quantum potential, more important to the
purpose here because of its dynamical implications, which is the fact
that this contribution to the quantum Hamilton-Jacobi equation arises from the kinetic
operator of Eq.~(\ref{eq2}).
Therefore, it is related to the diffusive part of the Schr\"odinger equation and not to
the action itself of an external potential function.
Therefore, even if it is used to explain in a sort of mechanistic way the
motion exhibited by quantum systems (within a Bohmian framework), it should be
interpreted as a kind of internal information conveyed to the system by its own
quantum state, which is continuously changing in time.
It is in this regard that the concept of ``mechanics'' is somehow dubious, because
the mechanism of the dynamical behaviors observed is partly due to the own system
(or, more strictly speaking, its quantum state).


\subsection{Dispersion of a localized Gaussian wave packet}
\label{sec24}

To illustrate the above facts in simple terms, consider the paradigm of localized quantum
system representing the free evolution of a particle of mass $m$, described by Gaussian
wave packet \cite{sanz:JPA:2008}.
In this case, although there are no
external forces acting on the particle, its states spreads out continuously in time,
first slowly and then, after undergoing an accelerating boost, linearly with time
\cite{sanz:AJP:2012}.
To better understand this behavior, consider that the particle is described by a normalized
one-dimensional wave packet centered at $x=0$,
\begin{equation}
 \Psi(x) = \left( \frac{1}{2\pi\sigma_0^2} \right)^{1/4} e^{-x^2/4\sigma_0^2} ,
 \label{eq8}
\end{equation}
with its width being $\sigma_0$.
The time-evolution of this wave packet is described by the time-dependent state
\begin{equation}
 \Psi(x,t) = \left( \frac{1}{2\pi\tilde{\sigma}_t^2} \right)^{1/4}
  e^{-x^2/4\sigma_0\tilde{\sigma}_t} ,
 \label{eq9}
\end{equation}
with
\begin{eqnarray}
 \tilde{\sigma}_t = \sigma_0 \left( 1 + \frac{\hbar t}{2m\sigma_0^2} \right) ,
 \label{eq10}
\end{eqnarray}
with its (time-dependent) width being
\begin{equation}
 \sigma_t = |\tilde{\sigma}_t| =
  \sigma_0 \sqrt{1 + \left( \frac{\hbar t}{2m\sigma_0^2} \right)^2} .
 \label{eq13}
\end{equation}
Because the initial momentum associated with the particle is zero, the center of this wave
packet remains at $x=0$ (there is no translational motion).
Yet, the expectation value of the energy is nonzero, but constant in time:
\begin{equation}
 \bar{E} = \langle \hat{H} \rangle = \frac{\hbar^2}{8m\sigma_0^2} .
 \label{eq11}
\end{equation}
This sort of average energy (\ref{eq11}) corresponds to the diffusive internal energy that
makes the wave packet to spread out with time, even if it can be recast as $\bar{E} = p_s^2/2m$,
in terms of the spreading momentum $p_s = \hbar/2\sigma_0$ \cite{sanz:JPA:2008}.
Nonetheless, we can still further investigate this contribution.
To that end, let us recast the wave packet (\ref{eq9}) in polar form and proceed with the
corresponding substitutions in Eq.~(\ref{eq3b}).
The kinetic contribution reads as
\begin{subequations}
\begin{equation}
 K = \frac{1}{2m}\left(\frac{\partial S}{\partial x}\right)^2 = \frac{\hbar^2}{8m\sigma_0^2}
  \left( \frac{\sigma_t^2 - \sigma_0^2}{\sigma_t^2} \right)
  \frac{x^2}{\sigma_t^2} ,
 \label{eq12a}
\end{equation}
while Bohm's quantum potential is
\begin{equation}
 Q = \frac{\hbar^2}{8m\sigma_0^2} \frac{\sigma_0^2}{\sigma_t^2}
  \left( 2 - \frac{x^2}{\sigma_t^2} \right) .
 \label{eq12b}
\end{equation}
 \label{eq12}
\end{subequations}

As it can be noticed from Eq.~(\ref{eq12a}), the lack of an initial momentum or the action
of an external potential makes more apparent how the quantum phase is responsible for the
generation of a dynamics, which eventually translates into the spreading of the wave packet.
This dynamics is null at $t=0$, but since the spreading factor, $\tilde{\sigma}_t$, in the
Gaussian state acquires a complex phase, it gradually induces the appearance of an internal
motion in the form of the spreading of the wave packet.
On the other hand, the quantum potential (\ref{eq12b}) is nonzero even at $t=0$, which is due
to its relationship with the curvature of the quantum state (either through its amplitude or
its density).
Because of their different origin, both quantities do not cancel each other or render
a constant value, but they keep evolving in time, spreading all over longer and longer
distances in configuration space, which explains the increasing spreading of free Gaussian
wave packets.

Following a standard Bohmian prescription and appealing to a typical description of dynamical
systems \cite{jordan-bk:1999}, it is seen that first the inverted parabola with $x$ 
corresponding to the quantum potential generates an unstable point (the kinetic term is zero),
which diverts trajectories towards positive and negative $x$ (with respect to $x=0$).
Then, the parabola describing the kinetic term starts gaining importance, which somehow helps
to counterbalance the action of the quantum potential, binding the motion.
Finally, at asymptotic times, the quantum potential becomes relatively flat compared to the
kinetic term (one term goes as $\sigma_t^{-2}$, while the other one goes as $\sigma_t^{-4}$),
which opens up gradually, as $x^2/t^2$, thus producing a linear spreading of the
trajectories (i.e., as $x/t$).

Nonetheless, their combined average, $\bar{K} + \bar{Q}$, when it is computed with respect
to the also time-dependent density $\rho$, i.e.,
\begin{equation}
 \bar{K} + \bar{Q} = \int_{-\infty}^\infty \left[ \frac{1}{2m}\left(\frac{\partial S}{\partial x}\right)^2
  - \frac{\hbar^2}{2m} \frac{1}{\rho^{1/2}} \frac{\partial^2 \rho^{1/2}}{\partial x^2} \right] \rho\ dx ,
 \label{eq14}
\end{equation}
renders the constant value (\ref{eq11}), which
is consistent with the fact that this type of motion is energy-preserving, as it is expected
from a free particle.
Actually, it is interesting to note that in the long-time limit, we obtain
\begin{equation}
 K + Q \approx \frac{1}{2} m \left( \frac{x}{t} \right)^2 ,
 \label{eq15}
\end{equation}
which is consistent with the asymptotic behavior displayed by $K$ in the long-time limit,
as it was commented above.
Noticed that Eq.~(\ref{eq15}) resembles the usual expression for the kinetic energy of a
classical particle (with $v = x/t$), although its average is (\ref{eq11}), as it can readily
be shown by averaging over $\rho$.
After all, in the long-time limit, the dispersion undergone by the Gaussian, Eq.~(\ref{eq13}), 
increases linearly with time as
\begin{equation}
 \sigma_t \approx v_s t ,
 \label{eq16}
\end{equation}
where $v_s = p_s/m$ \cite{sanz:JPA:2008}.
Actually, this expression not only provides the asymptotic spreading of a quantum wave
packet for a massive particle, but it also coincides with the divergence of a Gaussian
laser beam in paraxial optics \cite{sanz:ApplSci:2020} when the quantity $\hbar t/m$ is
substituted by $z/k$ (linear increase of the transverse dispersion with the longitudinal
coordinate $z$).

From a statistical viewpoint, the above discussion focuses around ensemble dynamics, which
is what $\rho$ eventually describes, namely the behavior of a large number of identically
distributed (non interacting) particles of mass $m$ in free space.
Following the standard Bohmian view (the one arising from his 1952 paper),
such particles follow well-defined trajectories, which are obtained after integrating
in time Eq.~(\ref{eq5}).
This is a legitimate way to understand such an equation.
However, there is an also legitimate but alternative way to understand it, namely, by
considering that such trajectories are just the streamlines that follow the flow described
by the local velocity field ${\bf v}$, as specified by Eq.~(\ref{eq5b}) and in compliance with
standard quantum mechanics.
In this case, irrespective of how real particles move (smoothly or randomly), Bohmian
trajectories only
reflect the local dynamics of the ensemble, just in the same way that a tiny floating particle
provides us with a clue on how a stream flows, but not on how each individual molecular
component of the stream behaves \cite{sanz:AJP:2012,sanz:JPhysConfSer:2012,sanz:FrontPhys:2019}.
This view is closer to Madelung's hydrodynamic formulation of the Schr\"odinger equation
\cite{madelung:ZPhys:1926}.
So, going to the point, after analytically integrating in time (\ref{eq5}) for the Gaussian
state (\ref{eq9}), the trajectories are found to follow the functional form 
\begin{equation}
 x(t) = \frac{\sigma_t}{\sigma_0}\ x(0) ,
 \label{eq17}
\end{equation}
where $x(0)$ is the corresponding initial condition.
As it can be noticed, these trajectories diverge in compliance with the divergence undergone
by $\sigma_t$, being the effect more apparent as their initial condition is chosen further
and further away from the center of the wave packet.
At each time, it is possible to determine how the distribution described by $\rho$
behaves by only inspecting the behavior exhibited by a swarm of such trajectories,
which provides us clues on different dynamical regimes \cite{sanz:AJP:2012}.


\section{Young's two slits revisited}
\label{sec3}

To better appreciate the implications of the Bohmian formulation of quantum
mechanics and their reach in our understanding of quantum phenomena, now we are going
to focus on the analysis of Young's two-slit experiment, which, quoting Feynman
\cite{feynman-bk1}, ``has in it the heart of quantum mechanics''.
As it is shown, when the phenomenology of this experiment is revisited in terms of Bohmian
mechanics, a different perspective arises on what is going on, which challenges its
traditional Copenhagian explanation.
Furthermore, because this experiment stresses in the simplest manner the capability of
quantum systems to exhibit delocalization while keeping long-distance space correlations 
(necessary to observe the well known interference fringes), by extension that analysis also 
provides us with a new physical understanding of the concept of coherence, central not only
to quantum mechanics, but also to optics.


\subsection{Current experiments, old explanations}
\label{sec31}

The single-event-based picture of interference rendered by Bohmian mechanics is, perhaps,
better understood by revisiting the experiment and how it has been traditionally explained.
As it was mentioned in the introductory section, there is a number of interference
experiments performed with both photons and material particles, which show that whenever
the particle flux is faint enough (to the point that each single detection can be monitored
in real time), a random-like distribution of detected events is observed instead of the
well-defined interference fringes obtained in a standard high-flux experiment.
In such experiments, interference fringes start emerging gradually from among the
point-like distribution of detected events as time passes by
\cite{padgett:AJP:2016,rose:AdvBiolMedPhys:1957}.
This can easily be illustrated by considering a simple experiment, which, in the current
case, has been performed by the author in the teaching optics laboratory (an experiment
that our students routinely perform every year).
It is a Young-type experiment performed with a 631~nm wavelength laser ($< 5$~mW power)
illuminating a thin steel mask with two parallel narrow slits.
The slits have an average width of 0.145~mm and their center-to-center distance is 0.865~mm.
The light coming from the slits is made to converge with a 150~mm focal length lens on a
CCD consisting of a $1024 \times 768$ array of square pixels (the side of each pixel is
4.64~$\mu$m long).

\begin{figure}[!t]
 \centering
 \includegraphics[width=\textwidth]{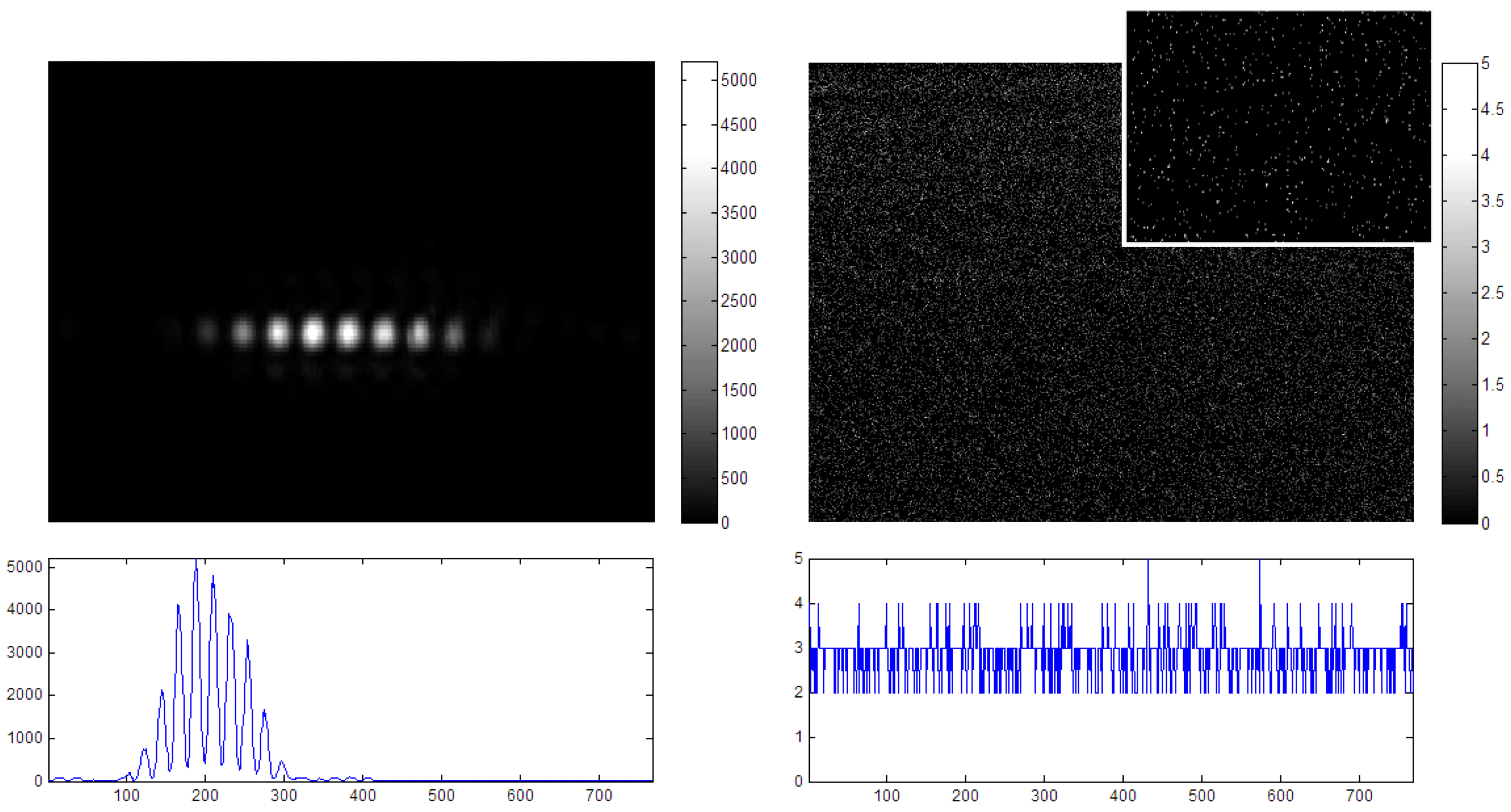}
 \caption{\label{fig1}
  Snapshots illustrating the light distribution produced by two narrow slits (0.145~mm
  wide) separated a distance of 0.865~mm when they are illuminated by a 631~nm wavelength
  laser under high-intensity conditions (left) and low-intensity conditions (right).
  The upper panels show the intensity distribution recorded by a CCD consisting of a
  $1024 \times 768$ array of square pixels (with a side 4.64~$\mu$m long).
  In both cases, the right-hand side gray-level scale in the upper panels denotes the
  intensity registered by each pixel during the time the experiment has been run (few
  seconds in both cases), which is proportional to the number of photons registered by
  each pixel.
  The lower panels show the intensity only along the transverse direction, that is,
  integrating (summing) over the vertical direction in the upper plot in order to make
  more apparent that the intensity can be assumed as a continuous distribution in the 
  high-intensity regime and as formed by discrete scintillations in the low-intensity
  regime.}
\end{figure}

The upper panels of Fig.~\ref{fig1} show two snapshots that illustrate the result of
performing the experiment under high-intensity conditions (left) and under low-intensity
conditions (right).
In both cases, a very short exposure time has been considered in order to emphasize who,
in the high-intensity regime, the light interference fringes appear as a continuous
intensity distribution.
On the contrary, when the intensity is low enough, which is achieved by placing two
polarizers in front of the laser source and then making their transmission axes to be
nearly perpendicular one another, we observe a series of uncorrelated scintillations
that distribute randomly across the detector surface.
By zooming in the upper right panel (see inset), we can notice that the distribution is
rather sparse, thus offering no clue on any underlying interference-type structure.
In order to get a better quantitative idea, the lower panels show the transverse
intensity distribution, which has been obtained by integrating (summing) the intensity of
the upper panels along the vertical direction.
The $x$-axis labels the position of the capture pixels along this direction, while the
$y$-axis provides us with the relative intensity, proportional to the number of photons
collected in the corresponding pixels (remember the summation over the vertical pixels
for a given $x$-position) during the experiment performance time (for a better read of
the upper panels, the gray-level scale to their right represents the same).
Regarding the interpretation of the data shown, several comments are in order.
First, while the horizontal axis in the lower panels runs over all the 768~pixels, for
a better visualization in the upper panels only the region around the fringes has been
considered (the same regarding the vertical axis).
Hence, the upper and lower panels cannot be directly compared, which is the reason for the
mismatching when comparing the maxima and minima in both cases.
Second, the discrete numbers that appear, in particular, in the lower right panel does
not correspond to photon counts or, in other words, to number of photons per pixel, but to
a quantity proportional to CCD units of counts.
Yet it is clear by comparing the two lower panels that while in one case the pattern runs
smoothly along the transverse direction (left), the same does not happen in the low-intensity
regime, where a nearly uniform discretized accumulation.
Finally, even with the limitations of not having at hand a reliable single-photon source, but
a simple setup (after all, the experiment has been carried out by a theoretician), it still
serves to the purpose of illustrating the discreteness involved in the formation of
interference patterns.
This phenomenon can only be noticed with a very faint illumination of the slits, but is of
much relevance in the understanding of the two-slit experiment.

\begin{figure}[!t]
 \centering
 \includegraphics[width=0.9\textwidth]{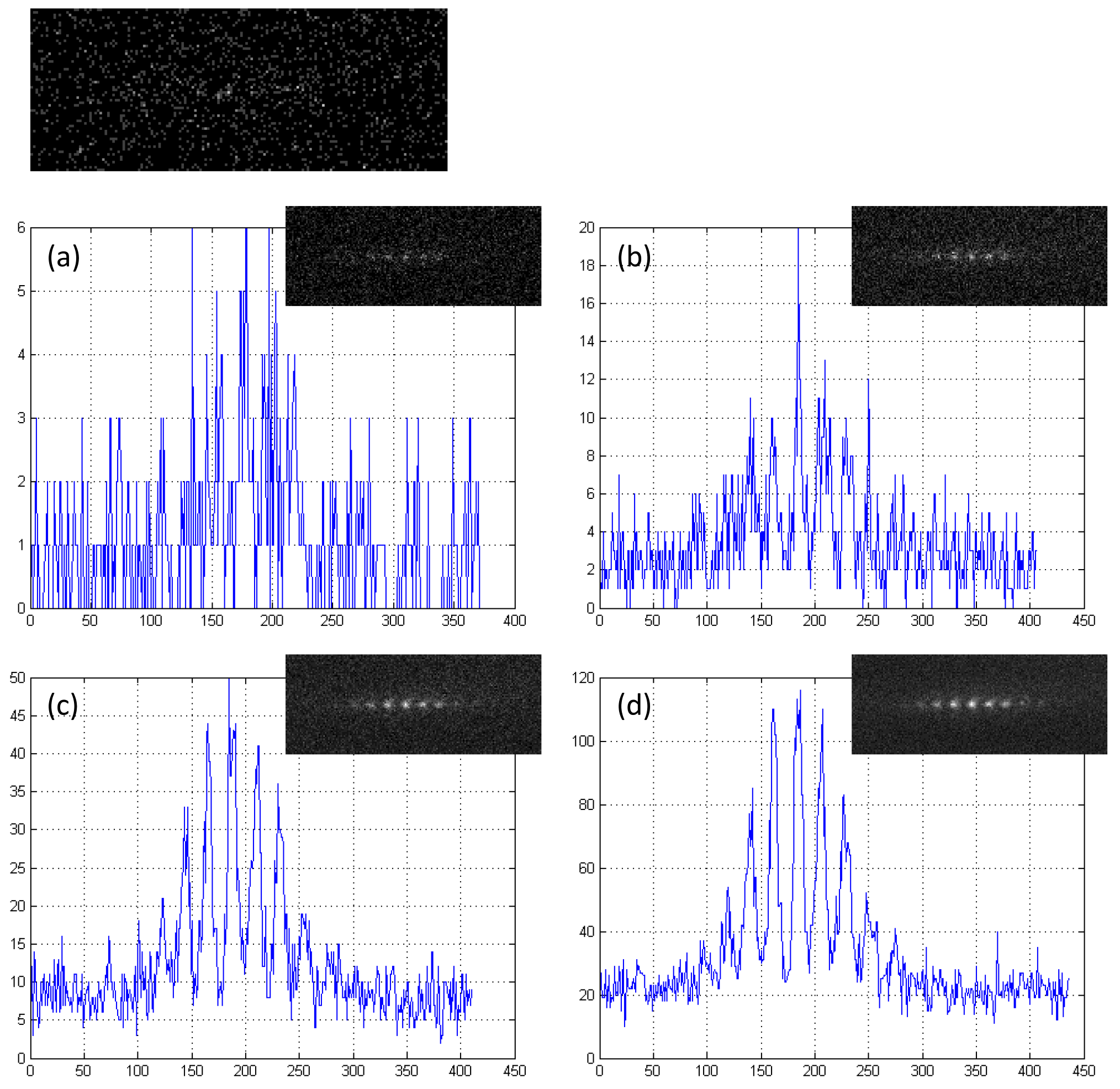}
 \caption{\label{fig2}
  From (a) to (d), snapshots taken at subsequent times for the experiment of Fig.~\ref{fig1}
  under low intensity conditions.
  The upper frame shows the random-type distribution of scintillations for a few seconds.
  Although there seems to be no correlation among those scintillations, as time proceeds it
  is seen that photons arrive in a larger proportion to certain pixels, while avoid others,
  thus giving rise to accumulations that increase beyond noise-type fluctuations.
  Eventually these accumulations give rise to an intensity distribution mimicking the one
  obtained under high-intensity conditions, which requires times of the order of several
  minutes (note that, even so, the maxima are close to relative intensity values of 120,
  while in the high-intensity regime (see Fig.~\ref{fig1}) they reached values up to 5,000).}
\end{figure}

In Fig.~\ref{fig2} the gradual appearance of the fringes is explicitly shown by beans of a
series of subsequent snapshots, each taken a larger exposure time.
In the upper panel of the figure there is a photograph for a short time, as in the right
panels of Fig.~\ref{fig1}.
As time proceeds, the sequence of fringes becomes more and more apparent, as it is shown in
the sequence from panels (a) to (d).
In this case, in order to focus on the region around the interference pattern, the intensity
in the 1D plots has been taken along a given range of the 2D photographs.
It can be noticed that these fringes are more apparent in the 2D plots than in the
representations of the relative intensity in terms of the transverse direction.
This is only an effect due to the summation over the vertical direction: since the pattern is
too faint, all other pixels exited with environmental noise photons are also going to contribute,
thus reducing the relative visibility of interference pattern.
This is the case in panels (a) and (b), for instance.
In the case of panels (c) and (d), the accumulation of photons in the regions around
interference maxima becomes more prominent than the background noise contribution.

The key question that arises now is that of the interpretation of the granular behavior involved
in the formation of interference fringes, as illustrated by the above experiment.
The traditional Copenhagian interpretation of the phenomenon is essentially based on the
pattern obtained in the high-intensity regime, that is, a continuous intensity distribution
that spreads all over some spatial region in the form of alternating light and dark spots
or fringes (regions with a high and low detection rates, respectively).
However, given that quantum systems consist of single, independent particles, the explanation
considers that, at some point the particle, understood a spatially localized system, becomes
and propagates as a wave before, while and after passing through the slits.
When it reaches the detector, the associated wave ``collapses'' and the particle acquires
again its corpuscular nature in the form of a spatially localized (single-event) detection
\cite{dempster:PhysRev:1927}.
Formally, this translates into the two different processes mentioned by Von Neumann to
describe the propagation and measurement of quantum systems \cite{vonNeumann-bk:1932}.
While the particle is not detected its evolution is unitary; when it is being detected, such
unitarity breaks down and it takes place a non-unitary irreversible ``collapse'' to a specific but previously indefinite spatial position.
This conception might seem odd and even uncomfortable, but it is what the experiment allows
us to know about the particle, even if we appeal to single-event experimental procedures, like
the one described above or all other that can be found in the literature; there is no way to go
further away and determine what is going on from the slits to the detector experimentally
without directly acting with the system, in which case interferences fly away.
However, it is also true that this is not the impression that one acquires when the particle
flux is weakened so much that the intensity distribution is finally reconstructed on an
event-to-event basis.
For some reason, many feel inclined to find a way to associated those individual, spatially
localized detections with the idea of a particle following a well-defined trajectory in space,
regardless of how this trajectory looks like, i.e., of which equation of motion describes it.
Therefore, even if cannot determine experimentally such trajectories, this ontic perspective
also seems to be a reasonable and legitimate explanation (neither better or worse than the
traditional Copenhagian collapse idea), which cannot be discarded (not, at least, with the
current experimental facts).

Apart from the single-event experiments mentioned so far, there are recent experimental
facts that are making us to reconsider our traditional conception of quantum systems
These changes are connected to the rather old concept of weak measurement
\cite{aharonov:PRL:1988,sudarshan:PRD:1989} and, more importantly, its experimental
implementation \cite{mir:NewJPhys:2007}.
Even though with its limitations regarding the interpretation of quantum phenomena (see below), this
technique has widened our view and understanding of quantum systems.
Contrary to a strong Von Neumann measurement (the usual measurement process in quantum mechanics),
a weak measurement only perturbs slightly the system, without making it to collapse, thus allowing
us to extract information on supplemental aspects of such a system at once \cite{lundeen:Nature:2011,kocsis:Science:2011}.
In the case of interference, both the probability density and the transverse flux, responsible for how
the former changes spatially with time [following Eq.~(\ref{eq4b})], can be determined within the same
experiment without requiring extra measurements, as it happens in quantum tomography, an also without
destroying the interference fringes.
With these two quantities, the local velocity field can be computed from Eq.~(\ref{eq5b})
at any time or, equivalently, any distance between slits and detector.
From here on, considering a series of initial conditions and integrating in time
(\ref{eq5}), one straightforwardly obtains the corresponding Bohmian trajectories,
as it is shown in \cite{kocsis:Science:2011} in the case of light (photons).

Of course, the information extracted from these experiments should be carefully considered,
avoiding conclusions that go beyond the experiment itself.
In the case we are dealing with, as it has been commented above (see Sec.~\ref{sec21}), there
is no empirical evidence on how to relate the inferred trajectories with the real motion of quantum particles.
These trajectories can be considered as streamlines accounting for the spatial dispersion of
ensembles, but not of individuals.
Note that although the average transverse flow, obtained from measurements over
many photons, behaves as specified by Eq.~(\ref{eq5b}) (or, to be more precise, in compliance
with the Maxwellian analog of this field \cite{sanz:AnnPhysPhoton:2010,sanz:ApplSci:2020}),
this does not mean that the detected photons follow Bohmian-type trajectories.
There are stochastic approaches, for instance, which also render the same average outcomes
\cite{bohm:pr:1954,bohm:PhysRep:1989,furth:ZPhys:1933,comisar:PhysRev:1965,nelson:pr:1966}.
Within this scenario, therefore, closer in spirit to Madelung's transformation of
Schr\"odinger's equation into a hydrodynamic form \cite{madelung:ZPhys:1926}, Bohmian
trajectories help us to understand the dynamics of this fluid when it reaches the two slits,
how it behaves after crossing them, giving rise to interference traits, or why interference
disappears if one of the slits is suddenly shut down (``observed'').
This brings in an alternative and very different picture of Young's two-slit experiment,
usually associated with the effects that follow the overlapping of the waves coming out from
each slit, and that has also overmagnified the role of the external observer in the removal
of the fringes.


\subsection{Dynamical role of the local velocity field}
\label{sec32}

Let us thus reconsider the problem from the Bohmian viewpoint.
To this end, a simple model based on the coherent superposition of two one-dimensional
Gaussian wave packets \cite{sanz:JPA:2008}.
In brief, to understand the model and its physical meaning, consider a screen with two slits
separated by a distance $d$ and both being parallel to the $y$-direction.
In this configuration, the $x$-axis cuts both slits in two symmetric halves and the $z$ axis
is perpendicular to the screen containing the slits.
If the slits are much wider along the $y$ direction than along the $x$ direction, and the
incident momentum, parallel to the $z$-axis, is relatively high, so that the angular spreading
by diffraction is negligible compared to the distance traveled along the $z$ axis, the wave
function of the system can be simplified by a product state, where interference takes place
along the (transverse) $x$ direction (further technical details on this modeling of diffraction
systems can be found in \cite{sanz:AOP:2015}).
To further simplify, the transmission function is assumed to be Gaussian, which produces two
diffracted Gaussian states at each slit.
In spite of its simplicity, this captures the essence of the phenomenon without any loss of
generality.
Accordingly, consider that the two diffracted waves are denoted by the coherent superposition
of two Gaussian wave packets,
\begin{equation}
 \Psi(x,t) = \psi_-(x,t) + \psi_+(x,t) ,
 \label{eq18}
\end{equation}
where each one of these wave packets has the same form as (\ref{eq9}) and subscripts
$\pm$ denote the position of their respective centers with respect to $x=0$, i.e., at
$x_\pm = \pm x_0$, with $x_0 = d/2$.
Recasting the wave packets in polar form, the following expressions for the probability
density and the quantum flux are readily obtained:
\begin{subequations}
\begin{eqnarray}
 \rho (x,t)\ & =\ & \rho_+(x,t) + \rho_-(x,t) + 2 \sqrt{\rho_+(x,t) \rho_-(x,t)} \cos \varphi(x,t) ,
 \label{eq19a} \\
 J(x,t)\ & =\ & \frac{1}{m} \Bigg\{ \rho_+(x,t)\ \frac{\partial S_+(x,t)}{\partial x}
  + \rho_-(x,t)\ \frac{\partial S_-(x,t)}{\partial x}
 \nonumber \\ & &
  \quad
  + \sqrt{\rho_+(x,t) \rho_-(x,t)}\ \frac{\partial \left[S_+(x,t) + S_-(x,t)\right]}{\partial x} \cos \varphi (x,t) \Bigg\}
 \nonumber \\ & &
  \quad
  + \frac{\hbar}{2m}\ \sqrt{\rho_+(x,t) \rho_-(x,t)}\
  \Bigg[ \frac{1}{\rho_+(x,t)}\frac{\partial \rho_+(x,t)}{\partial x}
 \nonumber \\ & & \qquad \qquad \qquad
  - \frac{1}{\rho_-(x,t)}\frac{\partial \rho_-(x,t)}{\partial x} \Bigg] \sin \varphi (x,t),
 \label{eq19b}
\end{eqnarray}
 \label{eq19}
\end{subequations}
with $\varphi (x,t) = [S_+(x,t) - S_-(x,t)]/\hbar$.
The superposition principle does not hold for any of these two magnitudes, since their
expressions involve the density and phase partial fields in a rather nonlinear fashion,
in particular, Eq.~(\ref{eq19b}).
From a dynamical point of view, this translates into an interesting property in the flux
that cannot be perceived in the probability density: at any time, it is zero at $x=0$.
This readily leads to an important physical consequence: the flux to the left of $x=0$
can never mix with the flux to the right \cite{sanz:JPA:2008}.
Accordingly, although the idea of constructive and destructive interference, based on
how the probability density is constructed, is formally correct, we find that the usual
explanation of the two-slits experiment is, to some extent, physically incorrect for it
neglects the dynamics of the probability density in terms of its flux.
When the latter is considered, the fact that the flows associated with each slit do not
mix implies that the left part of the interference pattern is always related to the left
slit, while the right part concerns to the right slit.
In other words, the flow makes distinguishable which part of the pattern is related with
each slit, even if there is no way to determine whether the same happens at an underlying
level with each individual real particle.

To above fact is better seen if Eqs.~(\ref{eq19}) are written explicitly in terms of the two
Gaussian wave packets and their parameters\footnote{For simplicity, the time-dependent
normalizing prefactor has been neglected, because it is dynamically irrelevant (it only 
induces the gradual decrease of both quantities as they
spread out spatially).}:
\begin{subequations}
\begin{eqnarray}
 \rho(x,t)\ & =\ & e^{-(x-x_+)^2/2\sigma_t^2} + e^{-(x-x_-)^2/2\sigma_t^2}
  + 2 e^{-(x^2 + x_0^2)/2\sigma_t^2} \cos (\kappa x) , \nonumber \\ & &
 \label{eq20a} \\
 J(x,t)\ & =\ & \frac{\hbar^2 t}{4m^2\sigma_0^2\sigma_t^2} \Big[ (x-x_+) e^{-(x-x_+)^2/2\sigma_t^2}
  + (x-x_-) e^{-(x-x_-)^2/2\sigma_t^2} \nonumber \\
 & & \qquad \qquad \quad + 2 x e^{-(x^2 + x_0^2)/2\sigma_t^2} \cos (\kappa x) \Big] \nonumber \\
 & & - \frac{\hbar x_0}{m\sigma_t^2}\ e^{-(x^2 + x_0^2)/2\sigma_t^2} \sin (\kappa x) ,
 \label{eq20b}
\end{eqnarray}
 \label{eq20}
\end{subequations}
with $\varphi = -\kappa x$ and $\kappa = \hbar t x_0/2m\sigma_0^2\sigma_t^2$.
Moreover, consider the timescale $\tau \equiv 2m\sigma_0^2/\hbar$, which provides us with a
measurement of the characteristic dispersion time associated with the wave packet and, hence,
different dynamical regimes characterizing its evolution \cite{sanz:JPA:2008,sanz:AJP:2012}.
As it can be noticed, for relative short times, $t \ll \tau$ ($\sigma_t \approx \sigma_0$),
when diffraction starts acting on each wave
packet but it is not enough to achieve their overlapping (an important value of $\rho$ in
the vicinity of $x=0$), Eq.~(\ref{eq20a}) describes two separate Gaussian distributions
\cite{sanz:AJP:2012,sanz-bk-2}.
In turn, the flux increases linearly with $x$ from negative to positive in both regions $x>0$
and $x<0$, where the sharp separation at $x=0$ removes any inconsistency.
This is a clear indication that, because both waves are present at the same time, there are
two dynamically separated spatial regions.

In the long-time limit, $t \gg \tau$ ($\sigma_t \approx \hbar t/2m\sigma_0$,
$\kappa \approx 2mx_0/\hbar t$), on the other hand, the probability density covers
long distances, $x \gg x_0$, and hence
\begin{subequations}
\begin{eqnarray}
 \rho(x,t)\ & \approx\ & 2 e^{-2m\sigma_0^2 x^2/\hbar^2 t^2}
  \Big[ \cosh (4m\sigma_0^2 x_0 x/\hbar^2 t^2) + \cos (2 m x_0 x/\hbar t) \Big] ,
 \nonumber \\ & &
 \label{eq21a} \\
 J(x,t)\ & \approx\ & \frac{2x}{t}\ e^{-2m\sigma_0^2 x^2/\hbar^2 t^2}
 \Big[ \cosh (4m\sigma_0^2 x_0 x/\hbar^2 t^2)
 + \cos (2 m x_0 x/\hbar t) \Big] \nonumber \\
 & & - \frac{2x_0}{t}\ e^{-2m\sigma_0^2 x^2/\hbar^2 t^2}
     \sinh (4m\sigma_0^2 x_0 x/\hbar^2 t^2) .
 \label{eq21b}
\end{eqnarray}
 \label{eq21}
\end{subequations}
The probability density, Eq.~(\ref{eq21a}), essentially consists of an oscillating function
modulated by a Gaussian prefactor, since the hyperbolic cosine grows spatially relatively slowly at a given time $t$ (like $x/t^2$, slower than the $x/t$ dependence of the Gaussian
prefactor or the cosine).
Accordingly, for a sufficiently large time, within the region of interest (ruled by the
argument of the Gaussian prefactor), the hyperbolic cosine can be assumed to be close to the unity, so that
Eqs.~(\ref{eq21}) can be conveniently further approximated as
\begin{subequations}
\begin{eqnarray}
 \rho(x,t)\ & \approx\ & 4 e^{-2m\sigma_0^2 x^2/\hbar^2 t^2} \cos^2 (m x_0 x/\hbar t) ,
 \label{eq21bba} \\
 J(x,t)\ & \approx\ & \frac{4x}{t}\ e^{-2m\sigma_0^2 x^2/\hbar^2 t^2}
   \cos^2 (m x_0 x/\hbar t) \nonumber \\
 & & - \frac{2x_0}{t}\ e^{-2m\sigma_0^2 x^2/\hbar^2 t^2}
     \sinh (4m\sigma_0^2 x_0 x/\hbar^2 t^2) .
 \label{eq21bbb}
\end{eqnarray}
 \label{eq21bb}
\end{subequations}
From Eq.~(\ref{eq21bba}) we notice that vanishing interference minima evolve in
time at a constant rate given by the expression
\begin{equation}
 v_\nu^{\rm min} = \left(2\nu + 1\right) \frac{\pi\hbar}{md} ,
 \label{eqcondmax}
\end{equation}
where $\nu = 0, \pm 1, \pm 2, \ldots$ denote the interferential order.
In turn, interference maxima (modulated by a Gaussian envelope) will also evolve
at the constant rate
\begin{equation}
 v_\nu^{\rm max} = \frac{2\nu\pi\hbar}{m d} .
 \label{eqcondmin}
\end{equation}
In both cases, the separation between adjacent interference minima and maxima remains
constant and depends on the inverse of the distance between the slits, as in the optical
Young experiment \cite{bornwolf-bk}.
Concerning Eq.~(\ref{eq21bbb}), despite the second term is negligible, it has not been
disregarded, because it plays a major role in the dynamics, as it will be seen later on.
Note that, whenever the probability density vanishes, this term does not.

\begin{figure}[!t]
 \centering
 \includegraphics[width=\textwidth]{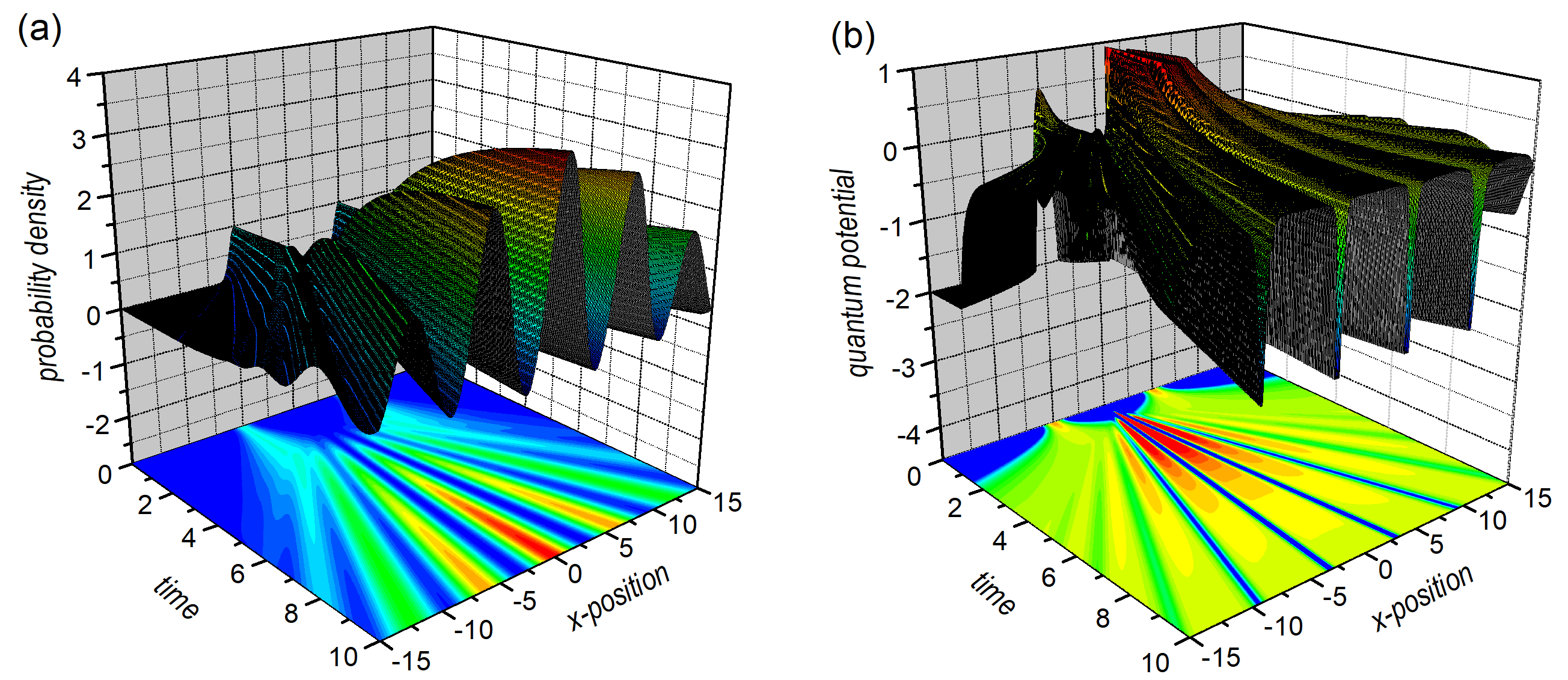}
 \caption{\label{fig3}
  Time-evolution of the probability density (a) and Bohm's quantum potential (b) for a
  coherent superposition of two Gaussian wave packets simulating Young's two-slit experiment
  along the transverse coordinate ($x$).
  In the color scale, lower values (nearly zero for the probability density and negative for
  the velocity field) are denoted with blue, while the higher ones are represented with red;
  note that, without any loss of generality, the quantum potential has been truncated both
  from the bottom and also the top due to the high positive and negative values that it
  reaches at some times and in some regions (in contrast with the nearly constant value that
  it acquires along the regions associated with the maxima of the probability density).
  The parameters considered in this simulation are $m = 1$, $\hbar = 1$, $\sigma_0 = 0.5$
  and $x_0 = 5$ ($d = 10$), in arbitrary units.}
\end{figure}

The two regimes can be seen in the numerical simulation displayed in Fig.~\ref{fig3}(a),
which represents the evolution of a coherent superposition of two time-evolving Gaussian
wave packets separated a distance $d=10$ ($x_\pm = \pm 5$) and with a initial width
$\sigma_0 = 0.5$ (without loss of generality, $\hbar = 1$ and $m = 1$).
These regimes can be better appreciated in the density plot below: two nearly freely
propagating Gaussians at the beginning, for $t \lesssim 1$, and well-defined interference
fringes for $t \gtrsim 4$, with their minima located at $x_\nu(t) = \pm 0.1\pi t, \pm 0.3\pi t, \ldots$ (note that in the particular case $t=10$, the maxima are located at
$x = \pm \pi, \pm 3\pi, \ldots$
From a standard Bohmian perspective \cite{dewdney:NuovoCimB:1979,holland-bk,sanz:prb:2000,sanz:JPCM:2002},
where the quantum potential is a central quantity, we can see, by inspecting
Fig.~\ref{fig3}(b), that there is not much difference, because it actually measures the
curvature of the probability density, following the second expression of Eq.~(\ref{eq6}).
Accordingly, the structure displayed by the quantum potential is going to be pretty similar
to that of the probability density, substituting the interference maxima of the latter by
plateaus and the nodes by deep minima or ``canyons'' (because of the analogy with these
geological formations), which also satisfy the condition (\ref{eqcondmin}).
Typically, it is assumed that, because the quantum potential is nearly flat between two
adjacent canyons (negligible quantum force, $-\nabla Q \approx 0$), particles are going
to accumulate in such regions, giving rise to the interference maxima, while they avoid
staying at such canyons, where they are affected by the action of an intense quantum force.
No doubt, the idea is appealing.
However, not only it provides redundant information with respect to the probability
density (as mentioned above, it measures its curvature), but totally neglects the dynamical
role of the quantum phase, necessary to explain, for instance, in the renowned non-crossing
property satisfied by the Bohmian trajectories.

In  order to provide a, say, non-redundant dynamical description to the emergence of the
interference pattern, let us get back to Eqs.~(\ref{eq21b}).
It consists of two contributions.
The first contribution is indeed the probability density multiplied by a prefactor $x/t$.
This prefactor is a (transverse) velocity that describes the overall spatial dispersion
(spreading) of the probability density.
Unlike the even parity displayed by the probability density (with respect to $x=0$), this
term has odd parity due to its additional dependence on $x$.
The second contribution, on the other hand, with an also odd parity, seems to play no role,
since it does not contain any information on interference and, moreover, decreases like
$t^{-1}$ all along $x$, thus becoming gradually less and less relevant.
Regarding the overall odd parity displayed by the flux, notice that it indicates that the
density is going to spread equally to both left and right.
Then, taking into account these features, how can the emergence of interference be explained
without relying again, as in the case of the quantum potential based explanation, on the 
probability density?

\begin{figure}[t]
 \centering
 \includegraphics[width=\textwidth]{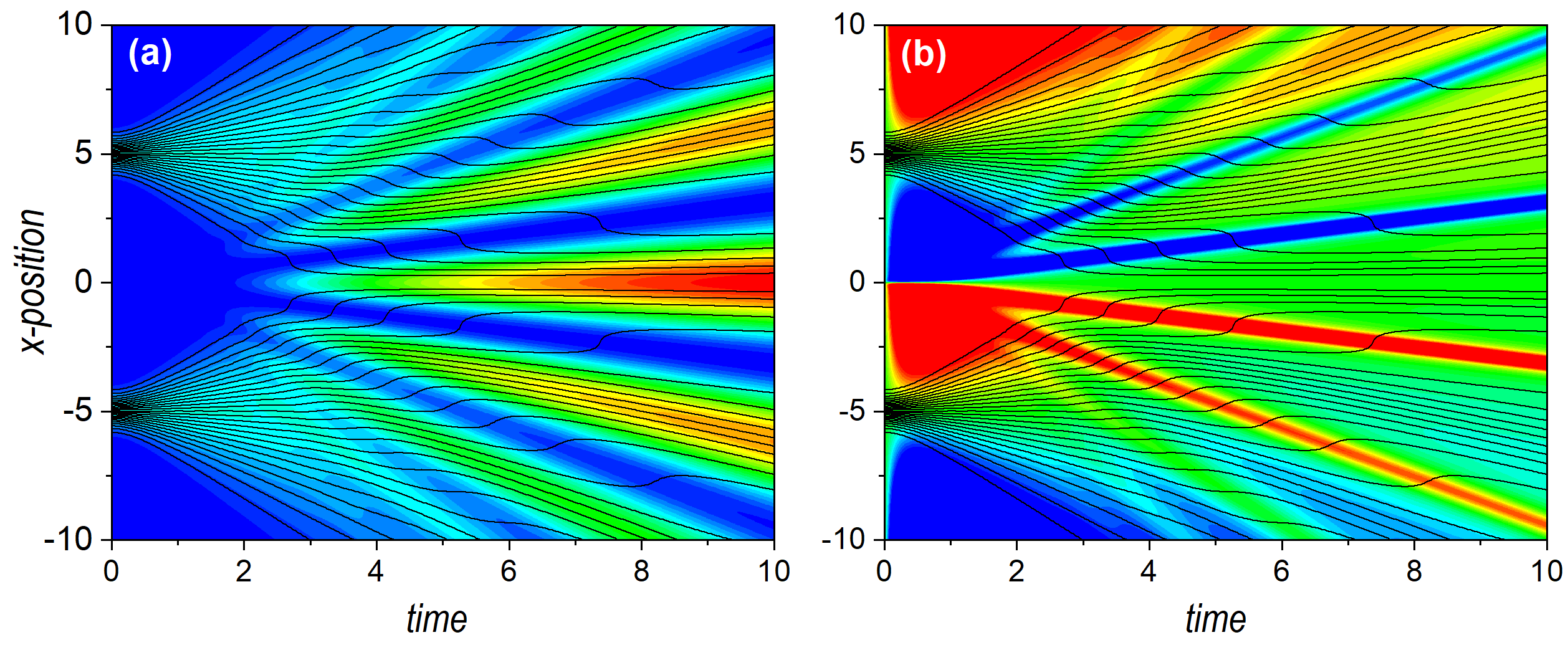}
 \caption{\label{fig4}
  Time-evolution of the probability density (a) and the local velocity field (b) for a
  coherent superposition of two Gaussian wave packets simulating Young's two-slit experiment
  along the transverse coordinate ($x$).
  In the color scale, lower values (nearly zero for the probability density and negative for
  the velocity field) are denoted with blue, while the higher ones are represented with red;
  in the case of the velocity field, vanishing values appear with green (along the central
  interference channel).
  For a better visualization of the dynamics, two sets of Bohmian trajectories, each one
  associated with one slit (Gaussian wave packet), are also on display (black solid lines).
  The parameters considered in this simulation are $m = 1$, $\hbar = 1$, $\sigma_0 = 0.5$
  and $x_0 = 5$ ($d = 10$), in arbitrary units.}
\end{figure}

To answer such a question, let us substitute Eqs.~(\ref{eq19}) into the second expression
of the equation of motion (\ref{eq5b}).
Thus, for the two wave packets, the latter equation reads as
\begin{eqnarray}
 \dot{x}(x,t)\ & =\ & \frac{1}{m}
 \Bigg\{ \frac{\rho_+(x,t)}{\rho(x,t)} \frac{\partial S_+(x,t)}{\partial x}
  + \frac{\rho_-(x,t)}{\rho(x,t)} \frac{\partial S_-(x,t)}{\partial x}
 \nonumber \\ & & \quad
  + \frac{\sqrt{\rho_+(x,t) \rho_-(x,t)}}{\rho(x,t)}\ \frac{\partial \left[S_+(x,t) + S_-(x,t)\right]}{\partial x}\ \cos \varphi(x,t) \Bigg\} \nonumber \\
 & & + \frac{\hbar}{m} \frac{\sqrt{\rho_+(x,t) \rho_-(x,t)}}{\rho(x,t)}
 \Bigg[ \frac{1}{\rho_+(x,t)}\frac{\partial \rho_+(x,t)}{\partial x}
 \nonumber \\ & & \qquad \qquad \qquad
  - \frac{1}{\rho_-(x,t)} \frac{\partial \rho_-(x,t)}{\partial x} \Bigg] \sin \varphi(x,t) ,
 \label{eq23}
\end{eqnarray}
where $\rho(x,t)$ is as given by Eq.~(\ref{eq19a}).
If we now reconsider the above limits, we find that, for short times, although the
probability density concentrates around either $x_+$ or $x_-$, Bohmian trajectories
associated with each wave packet are going to evolve seemingly like if the other wave
packet has no influence on them, i.e., like the trajectories related to a single Gaussian
wave packet problem \cite{sanz:AJP:2012,sanz:cpl:2007}.
This situation can be seen in Fig.~\ref{fig4}(a) up to $t \simeq 1$ (beyond this time the
trajectories closer to $x=0$ start undergoing deviations from the single wave-packet case).
However, this is only in appearance, as it can be noticed by inspecting Fig.~\ref{fig4}(b),
where the density plot represents the associated local velocity field, which changes very
abruptly at $x=0$, as expected according to the above discussion based on the quantum flux.
It is observed that, within the spatial domain of each slit, the flux associated to each
slit is the same and, therefore, the corresponding trajectories are expected to display
the same behavior.

As time increases and the probability density starts exhibiting interference maxima and
minima [see Fig.~\ref{fig4}(b)], the two swarms of Bohmian trajectories acquire a non-regular
distribution, loosing information about each particular slit and evolving along the
directions indicated by the maxima, while they undergo dramatic turns at nodal regions in
order to avoid them.
This is in agreement with the above quantum potential based argumentation.
But, why does this happen?
If Eqs.~(\ref{eq21bb}) are explicitly substituted into Eq.~(\ref{eq5b}), i.e., if the
long-time limit is considered in Eq.~(\ref{eq23}), this latter equation can no longer be
reduced to either one wave packet or the other, but needs to consider the full wave.
The equation of motion (\ref{eq5b}) reads now as
\begin{equation}
 \dot{x} \approx \frac{x}{t} - \frac{x_0}{2t}\
  \frac{\sinh (4m\sigma_0^2 x_0 x/\hbar^2 t^2)}{\cos^2 (m x_0 x/\hbar t)} .
 \label{eq24}
\end{equation}
Following this expression for the local velocity it is clear that, out of the reach of the
nodes and at a given time, the flux increases linearly with the position, since the first
term on the r.h.s.\ is the dominant one.
This simply means that trajectories will move apart from $x=0$ either with positive velocity
in the positive half-plane or with negative velocity in the negative half-plane, as it is
observed in the case of a simple Gaussian wave packet \cite{sanz:AJP:2012}.
Actually, if we take into account that the maximum value of the cosine in the second term
becomes maximum for $m x_0 x/\hbar t = \nu \pi$, with $\nu = 0, \pm 1, \pm 2, \ldots$, on
average, the expression for the velocity between two neighboring nodes is given by
\begin{equation}
 \dot{x} = \frac{\nu \pi\hbar}{m x_0} = \bar{v} \nu ,
 \label{eq24b}
\end{equation}
where $\bar{v} \equiv \pi\hbar/m x_0 = 2\pi\hbar/md$.
This means that the velocity is a quantized quantity, showing a ladder-type structure, where
each step has nearly the same width and describes an interference channel, i.e., a region that
will accommodate an interference maximum of the probability density.
These structures are typical whenever diffraction channels appear regardless of whether they
have been produced by two slits \cite{luis:AOP:2015}, many slits \cite{sanz:AOP:2015} or
scattering with a metal surface \cite{sanz:EPL:2001}.

However, at the nodes, the numerator of the second term cancels out and the velocity acquires
a sudden change or kick either below or above the value indicated by the first term.
If the particle is in the positive half-plane, the kick is negative; if its in the negative
half-plane, then the kick is positive.
This translates into a reorientation of the trajectories, which instead of being pulled
apart, as in the case of the Gaussian wave packet, they are gradually redirected towards
inner interferential maxima.
This can be seen in Fig.~\ref{fig4}(b), where the effect of the kicks appears as a fast
twist in the trajectories each time they approach a nodal point, inducing the passage of the
trajectories from the region associated with an interference maximum to the immediately
nearby one (that is, the motion takes place in discrete jumps, one by one).
In the present example, this implies that central maxima eventually become more populated
than marginal ones, which, in turn, serves us to understand the evolution of the probability
density displayed in Fig.~\ref{fig3}(a).
This behavior, though, can readily be generalized to grating diffraction, thus providing an
extremely natural interpretation for the appearance of diffraction orders, Bragg's law or
the relationship between the beam size and the definition of diffractive features
\cite{sanz:AOP:2015}.
Analogously, the same can be directly translated to the realm of optics,
providing a more intuitive picture of interference and diffraction phenomena
\cite{sanz:AnnPhysPhoton:2010,sanz:JRLR:2010,luis:AOP:2015,sanz:ApplSci:2020}, which is
excellent agreement with the experimental findings reported by Steinberg and coworkers
regarding Young's two-slit experiment several years ago \cite{kocsis:Science:2011,sanz:PhysScr:2013,sanz:EPN:2013}.

Summing up, it is interesting to note that, although the standard or traditional Bohmian
view based on Bohm's quantum potential does not allow to explain the dynamical origin of
the renowned non-crossing rule among trajectories, the local velocity field provides a
rather complete understanding of the phenomenon.
Again, this does not mean that quantum particles cannot move in the most unexpected manners,
because that is, so far, a challenging unknown.
It simply means that the flux describing the evolution of the probability density,
which describes how the swarm of quantum particles behaves (distributes) spatially on average
at each time, exhibits a very precise dynamics, according to which it is indeed possible, to
some extent, to establish a well-defined separation between the regions covered by each slit
at a dynamical level.
None of the (tracer) Bohmian trajectories starting in one of the slits will ever cross the
region dominated by the other slit, and vice versa.
This is, therefore, a physical manifestation (or evidence) of the quantum phenomenon or
quantum resource that we call coherence.
As a consequence, two-slit experiments turn out to be equivalent to single slit experiments
coupled to short-range attractive walls \cite{sanz:JPA:2008}, where the presence of the
attractive well induces the appearance of long-living resonances near the wall (associated
with half the central maximum).
This picture is quite far from the usual one, although it resembles the typical reduction
in two-body classical scattering problems, where the two systems are substituted by a single
one acted by an effective central force.

Furthermore, there is another important related consequence.
By inspecting Eq.~(\ref{eq23}), it is noticed that in order to remove any trait of coherence,
not only the interference term must be somehow removed, as it is usually
mentioned when dealing with the removal of interference in the two-slit experiment.
The disappearance of such a term simply means that the flux does not
include any wavy term.
However, the information about the existence of two slits open at once still persists.
Therefore, the non-crossing rule is still preserved, i.e., the Bohmian trajectories leaving
each slit are not going to mix \cite{sanz:EPJD:2007,luis:AOP:2015}.
The fact that both slits still influence the dynamics means that there is coherence even if
there is no interference.
In order to remove all traits of coherence, it is important to also remove information about
the other slit \cite{sanz:CPL:2009-2}.
In the traditional picture of the two-slit experiment this actually happens when we decide
to include the action of an external observer (detector), which just removes the contribution
of one of the slits.
Within the more refined description provided by the theory of open quantum systems
\cite{breuer-bk:2002}, this removal is simply the effect of the different manner that the
system gets entangled with an environment when it crosses one slit or the other
\cite{sanz:FrontPhys:submit21}.


\section{Final remarks}
\label{sec4}

Quantum mechanics is supposed to be the mechanical theory of quanta, that is, of the
`bits' of matter (electrons, atoms, molecules, etc.) and radiation (photons).
However, what Bohm put forth was that quantum mechanics had more of a non-mechanical
theory, because of the important role played by the whole over the individual.
Formerly, it was proposed as a counterexample to Von Neumann's theorem on the impossibility
of hidden variables in quantum mechanics, although the fact that it introduces into this
theory a language pretty similar to that of the classical Hamilton-Jacobi formulation has
led to associate the corresponding trajectories with the actual motion displayed by real
quantum particles.
Thus, in the same way that a classical (interaction) potential function determines the
motion of a (classical) particle, in the quantum mechanical case it would be the combined
action of such a function plus the so-called Bohm's quantum potential the mechanism behind
the topology displayed by the trajectories pursued by quantum particles.
Of course, this potential acts on quantum particles even in the case of free motion,
where there is no external interaction ($V=0$).
In such a case, the bare action of Bohm's potential shows very nicely how it accounts
for pure quantum effects, such as interference, as it can be seen not only with a free
particle (a freely released wave packet), but also in slit diffraction problems
\cite{sanz:JPCM:2002}.
Much has been discussed in the literature about this potential, its properties and its
applications (see \cite{holland-bk} and references therein, for instance), yet it is nothing
but a measure of the local instantaneous curvature of the probability density, thus providing
us to some extent with redundant  information (the same information already provided by the
probability density).
Nonetheless, recently it has received some attention as a magnitude that, in principle,
could be measured, particularly if instead of massive particles one considers light
\cite{umul:Optik:2020,hojman:arxiv:2021}, taking advantage of the one-to-one correspondence
between Schr\"odinger's equation and the paraxial Helmholtz equation.
Of course, this is not impossible, as it has also been the case of the transverse momentum
\cite{kocsis:Science:2011}, even though stricto sensu none of these quantities correspond to
quantum observables.

In order to avoid such redundancy, here the discussion has turned around the phase field
and, more specifically, the associated local velocity field, which allows us to establish a
connection between the probability density and the quantum flux, thus avoiding the extra
Bohmian postulate of a guidance condition.
The implications of the single valuedness of the quantum phase have long been discussed in
the literature to explain the non-crossing property exhibited by Bohmian trajectories in
the configurations space \cite{holland-bk}.
Unfortunately, the quantum phase only manifests through interference, thus providing little
clue on the dynamics displayed by the probability density.
The local velocity field, in turn, is a well-defined quantity with a precise physical
meaning, which can be experimentally determined through weak measurements, as shown in
\cite{kocsis:Science:2011}.
Accordingly, we have analyzed the dynamical information rendered by this quantity, which
allows us to understand and explain the time-evolution shown by the probability density at
each point of the configuration (in positions) space.
In analogy to classical hydrodynamic systems, this velocity field can be probed by
launching a series of tracers and let them to move accordingly, which provides us with
a more precise picture at a local level of the probability flux across the configuration
space in the form of probability-flow streamlines or trajectories.
These trajectories are the usual Bohmian trajectories, which here arise in a natural way,
without any need to introduce the concept of hidden variable.
It is in this way, used as tracers of the quantum dynamics, that Bohmian trajectories
constitute a remarkably beneficial tool to probe and understand quantum phenomena with
a language (that of dynamical systems) closer to our experience than abstract Hilbert
algebras ---perhaps more appropriate from a formal viewpoint, but totally useless to
understand what is going on in a real-lab experiment, where we know that we have something
that goes from somewhere to somewhere else, which can be acted and measured, etc.

With the purpose to illustrate the advantages of the local velocity field as a convenient
tool to analyze and explain quantum dynamics, Young-type interference has been studied.
Thus, while the usual Bohm's potential view provides an interpretation similar to the
Newtonian one (particles moving in regions with nearly constant potential values, while
avoiding others with strong, sudden changes), the velocity field provides us with a more
precise description of different dynamical regions and regimes.
Accordingly, it is seen that, the center of symmetry of the system, namely, the axis $x=0$
in our case, is a zero-flux line, which divides the configuration space into two dynamically
different regions.
Of course, the fact that the flux vanishes along this axis, and so the velocity field,
does not preclude the possibility that, at a ``subquantum'' level (i.e., at a level below the
equilibrium described by Schr\"odinger's equation), particles might randomly cross this
axis from one region to the other and vice versa, as it happens with chemicals (reactants
and products) under dynamical equilibrium conditions in a chemical reaction.
Yet this makes an important difference with respect to the traditional explanation
attributed to the two-slits experiments, where there is total indistinguishability.
Here, the distinguishability of dynamical domains gives rise to trajectories leaving one of
the slits that ``know'' of the existence of trajectories leaving the other slit.
These trajectories probe the dynamics associated with the flux, thus providing no clue on
how the motion of real particles might be, but only the resulting average (equilibrium)
motion.
The reveal how the velocity field changes locally at each time, undergoing fast and sudden
turns whenever there is a strong variation (similar to kicks), while moving nearly parallel
in those regions with (nearly) constant velocity.
The latter region happen to be quantized, i.e., the average velocity changes in units of
$\pi\hbar/m x_0 = 2\pi\hbar/md$ from one to the immediately neighboring one (both separated
by a kick).
Each one of these regions constitutes an interference channel, i.e., a region along which
trajectories tend to keep moving in a Newtonian sense.
These highly populated regions correspond to the interference maxima displayed by the
probability density.

If the mechanism to avoid regions with strong variations of the velocity field is clear,
which makes trajectories to get promoted from the outer interference channels to the
innermost ones, the same does not hold to explain why trajectories cannot cross the
$x=0$ axis, along which trajectories coming from both slits align.
The non-crossing here arises from having two different dynamical regimes well defined since
the very beginning.
Note here another deviation with respect to the standard view in terms of a simple
superposition relation, which holds true formally, but that cannot be accepted in dynamical
terms, since it will only appear provided both slits (both diffracted waves) are present
since the very beginning.
Accordingly, the concept of coherence acquires a different but totally unambiguous physical
meaning, in terms of the equivalence between this problem and that one of a single particle
colliding with an attractive potential wall \cite{sanz:JPA:2008}.
The attractive well happens to be relatively shallow, but with an extension beyond the two
wave packets (diffracted beams), which implies their mutual knowledge even if the probability
density is negligible in between.
Only if the information about the existence (presence) of one of the slits disappears (either
gradually or suddenly), trajectories from one domain will start crossing the trajectories from
the other domain \cite{luis:AOP:2015,sanz:EPJD:2007,sanz:CPL:2009-2}.
This situation thus describes a (partial or total) loss of coherence, which happens when
the system is strongly entangled with another environmental subsystem \cite{sanz:FrontPhys:submit21}.

At this point, based on the above discussion, one may still wonder whether real quantum
motion is still accessible.
As mentioned above, if it exists, it must be found at a subquantum level.
Nonetheless, the fact that the local value of the velocity field (transverse momentum) can be
experimentally determined opens new perspectives in our understanding of the quantum world.
Now we know that not only quantum particles distribute according to the usual probability
density even though they are totally uncorrelated, but also that in order to do it they
necessarily form currents.
This means that the usual Copenhagian view in terms of particle becoming a wave during the
experiment and then a particle again at the detector is currently getting blurred.
We have a precise quantum dynamical description of the average (equilibrium) behavior of
quantum systems where both their distribution (probability density) and spatial motion
(velocity field) can be determined without violating any of the fundamental principles
of quantum mechanics.

Because of the empirical impossibility to relate Bohmian trajectories with the actual
paths followed by real particles (in the sense that no experiment will be able to reveal
this very motion), one might wonder whether it provides or not a solution to the so-called
measurement problem \cite{zurek-bk}.
It is clear that, apart from point-like particles traveling along well-defined trajectories,
a proper description of such a problem requires including explicitly the presence of a second
agent or system, namely the detector.
When doing so, entanglement immediately arises \cite{giulini-bk,schlosshauer-bk:2007},
which is the physical mechanism behind the fact that we observe the system ``collapsing''
on any of the detector pointer states.
However, if we consider the simple experiment presented here, this is eventually equivalent to
make statistics over arrivals at certain spatial regions (pixels of a given finite dimension).
At this level, there is no need, therefore, to provide a better description of how each
photodetector state acts on or gets entangled with the system wave function, because each arrival
itself can be counted (registered).
The collection of these arrivals over time will provide us with a relatively fair picture of the
detection at a local level (pixel by pixel), which is equivalent to monitor in time the formation
of a full image over the whole scanning surface (e.g., a two-slit interference pattern with photons
or electrons, or, in the case of incoherent light, the appearance of a photograph).
In the standard quantum-mechanical approach the same is not possible, because the wave function
considered gives us the full solution (full image), even if later on some treatments (convolving
functions) are required in order to adapt such a solution to the finite-sized detection elements
(pixels, slits, etc.).
In this sense, and leaving aside other ontological connotations, a fully quantum-mechanical
trajectory-based approach proves to be more powerful than other standard quantum approaches,
since we are able to obtain first-principle theoretical descriptions of quantum phenomena
closer to real-life experiments without the need of extra treatments or elements (only the
necessary ones).
So, in conclusion, once the ``mysticism'' that usually accompanies Bohmian mechanics is removed,
whether this first-principle view can be considered of potential interest at a computational or
a fundamental level it is left to the reader's opinion.


\section*{Acknowledgements}

This work is partly based on the opening talk of the Journ\'ees Louis de Broglie
(November, 2019).
The author is much grateful to Thomas Durt and all the Organizers of the Journ\'ees for their
kind invitation to participate in this important meeting for those who think that other
understandings of quantum mechanics are possible.
He is also indebted to the Fondation Louis de Broglie for the kind attention received from
this institution.

Support to produce this work is also acknowledged to the Spanish Research Agency (AEI)
and the European Regional Development Fund (ERDF), under Grant No.\ FIS2016-76110-P.





\end{document}